# Near Infrared Microspectroscopy, Fluorescence Microspectroscopy, Infrared Chemical Imaging and High-Resolution Nuclear Magnetic Resonance Analysis of Soybean Seeds, Somatic Embryos and Single Cells

I.C. Baianu[*1-3], D. Costescu[2-3], T. You[1-2], P.R. Lozano[1-2], N.E. Hofmann[2] and S.S. Korban[4]

[1]FSHN Dept., [2]Agricultural Microspectroscopy NIR & NMR Facility, [3]Nuclear, Plasma and Radiological Engineering Dept., and [4]Department of Natural Resources & Environmental Sciences, ACES College, University of Illinois at Urbana-Champaign, Urbana, Illinois 61801, USA

**Abstract**

Novel methodologies are currently being developed and established for the chemical analysis of soybean seeds, embryos and single cells by Fourier Transform Infrared (FT-IR), Fourier Transform Near Infrared (FT-NIR) Microspectroscopy, Fluorescence and High-Resolution NMR (HR-NMR). The first FT-NIR chemical images of biological systems approaching 1micron (1μ) resolution are presented here. Chemical images obtained by FT-NIR and FT-IR Microspectroscopy are presented for oil in soybean seeds and somatic embryos under physiological conditions. FT-NIR spectra of oil and proteins were obtained for volumes as small as $2\mu^3$. Related, HR-NMR analyses of oil contents in somatic embryos are also presented here with nanoliter precision. Such 400 MHz [1]H NMR analyses allowed the selection of mutagenized embryos with higher oil content (e.g. ~20%) compared to non-mutagenized control embryos. Moreover, developmental changes in single soybean seeds and/or somatic embryos may be monitored by FT-NIR with a precision approaching the *picogram* level. Indeed, detailed chemical analyses of oils and phytochemicals are now becoming possible by FT-NIR Chemical Imaging/ Microspectroscopy of single cells. The cost, speed and analytical requirements of plant breeding and genetic selection programs are fully satisfied by FT-NIR spectroscopy and Microspectroscopy for soybeans and soybean embryos. FT-NIR Microspectroscopy and Chemical Imaging are also shown to be potentially important in functional Genomics and Proteomics research through the rapid and accurate detection of high-content microarrays (HCMA). Multi-photon (MP), pulsed femtosecond laser NIR Fluorescence Excitation techniques were shown to be capable of Single Molecule Detection (SMD). Therefore, such powerful techniques allow for the most sensitive and reliable quantitative analyses to be carried out both *in vitro* and *in vivo*. Thus, MP NIR excitation for Fluorescence Correlation Spectroscopy (FCS) allows not only *single molecule detection*, but also molecular dynamics and high resolution, *submicron* imaging of *femtoliter* volumes inside living cells and tissues. Thesenovel, ultra-sensitive and rapid NIR/FCS analyses have numerous applications in important research areas, such as: agricultural biotechnology, food safety, pharmacology, medical research and clinical diagnosis of viral diseases and cancers.

______________________________________________
*Corresponding Author: Professor I.C. Baianu

**KEYWORDS:**

FT-NIR and FT-IR Instruments, applications of FCS/NIR, Agricultural biotechnology, IR Chemical Imaging and NMR, Microspectroscopy, DNA/RNA Micro-array analysis by NIR, High resolution and super-resolution FT-NIR/IR, IR Chemical Imaging by FPAW, Spotlight 300 Microspectrometer, Two photon NIR excitation for FCS, Single Cell and sister 1-molcule dynamics, FCS of molecules, single cells, Soybean oil, protein and moisture analysis, FT-NIR and FT-IR, high-resolution NMR of soybean oil in seeds and somatic embryos, chemical mutagenesis of soybean embryos, picomole FT-NIR and femtomole FCS-NIR analysis of single cells, phytochemicals detection in soybean seeds and cells by FT-NIR, high-power, femtosecond Ti:Sapphire NIR excitation for FCS, FCS/PCR, Nucleic acid hybridization, FT-IR and FT-NIR Images of Soybeans and Embryos, FT-IR and NIR Chemical Imaging Tests, Spatial Resolution Test, FT-NIR Micro-Imaging, FT-NIR Images of Soybeans and Embryos, FT-IR Reflectance Chemical Images, Somatic Embryo, NIR Reflectance Chemical Image of a Red Coat Azuki Red Bean, FT-NIR Chemical Imaging by Difference Spectroscopy (CIDS), High Resolution NMR Analysis of Soybean Oil in Somatic Embryos, HR NMR, $^1$H NMR Spectrum of the somatic embryogenic culture of a soybean sample, TEM Micrograph of a Suspension of Soybean Somatic Embryos in Culture, Single Molecule Detection, two-photon excitation, one-photon excitation, three-photon-excitation, Fluorescence Correlation Spectroscopy (FCS), Fluorescence Resonance Energy Transfer (FRET), Fluorescence Lifetime Imaging Microscopy (FLIM), Fluorescence Recovery After Photobleaching (FRAP), Single Photon Confocal Fluorescence Correlation Spectroscopy, Inverted Epifluorescence Microscope, FCS auto-correlation, Fluorescence Fluctuations, Fluorescence Intensity, Fluorescence Correlation Spectroscopy and Imaging Experiments in Solutions and Plant Cell Suspensions, Pulsed, Two-Photon NIR Laser Excitation, Multi-photon (MPE) NIR excitation, FCS Alba Spectrometer Microspectrometer System, FCCS Cross-Correlation with Two Fluorescent Labels, FCCS Applications to DNA Hybridization.

**TABLE OF CONTENTS**

# 1. INTRODUCTION

Infrared (IR) and Near Infrared (NIR) commercial spectrometers employ, respectively, electromagnetic radiation in the range from to ~150 to 4,000 cm$^{-1}$, and from 4,000 to ~14,000 cm-$^{1}$. The utilization of such instruments is based on the proportionality of IR and NIR specific absorption bands with the concentration of the molecular components present, such as protein, oil, sugars and/or moisture. The molecular bond stretching/vibrations, bending and or rotations cause specific absorption peaks or bands, centered at certain characteristic IR and NIR wavelengths. FT-IR/NIR spectrometers obtain spectra using an interferometer and also utilize Fourier Transformation in order to convert the interferogram from the time domain to the frequency domain. The use of interferometry in FT-IR and FT-NIR spectroscopy increases the spectral resolution, the speed of acquisition, the reproducibility of the spectra and the signal to noise ratio in comparison with dispersive instruments that utilize either prisms or diffraction gratings.

An FT-IR/NIR image is built up from hundreds, or even thousands, of FT-IR/NIR spectra and is usually presented on a monitor screen as a cross-section that represents spectral intensity as a pseudo-color for every microscopic point in the focal plane of the sample. Special, 3D surface projection algorithms can also be employed to provide more realistic representations of microscopic FT-IR/NIR images. Each pixel of such a chemical image represents an individual spectrum and the pseudo-color intensity codes regions with significantly different IR absorption intensities. In 2002, four commercial FT-IR/NIR instruments became available from PerkinElmer Co. (Shelton, CT, USA): an FT-NIR Spectrometer (*SpectrumOne-NTS*), an FT-NIR Microspectrometer (*NIR AutoImage*), an FT-IR Spectrometer (*SpectrumOne)* and an FT-IR Microspectrometer (*Spotlight300*). The results of the tests obtained using these four instruments are shown in section 3.1.

The employment of high-power, pulsed NIR lasers for visible fluorescence excitation has resulted in a remarkable increase of spatial resolution in microscopic images of live cells, well beyond that available with the best commercial FT-NIR/IR microspectrometers, allowing even for the detection of single molecules. This happens because fluorescent molecules can absorb two NIR photons simultaneously before emitting visible light, a process referred to as "*two-photon excitation*." Using two-photon NIR excitation (2PE) in a conventional microscope provides several great advantages for studying biological samples. As the excitation wavelength is typically in the NIR region, these advantages include efficient background rejection, very low light scattering and low photodamage of unfixed biological samples and *in vivo* observation. Additionally, photobleaching is greatly reduced by employing 2PE, and even more so in the case of three-photon NIR excitation (3PE). The spatial region where the 2PE process occurs is very small (of the order of 1 femtoliter, or 10$^{-15}$ L), and it decreases even further for 3PE. Multiphoton NIR excitation allows submicron resolution to be obtained along the focusing (z) axis in epi-fluoresence images of biological samples, without the need to employ any confocal pinholes. The 2PE and 3PE systems with ~150-femtosecond (10$^{-13}$ s) NIR pulses have several important advantages in addition to high resolution. Firstly, they offer very high sensitivity detection of nanomole to femtomole concentrations of appropriately selected fluorochromes. Secondly, these systems have very high selectivity and the ability to detect interactions between pairs of distinctly fluorescing molecules for intermolecular distances as short as 10 nm, or less. 2PE and 3PE also allow one to rapidly detect even single molecules through Fluorescence Correlation Spectroscopy (FCS); FCS is usually combined with microscopic imaging. The principles of single photon FCS microscopy are briefly discussed next, in Section 2.2.

## 2. PRINCIPLES

A complete understanding of the principles of chemical imaging as well as fluorescence microscopy that allow the quantitative analysis of biological samples is necessary in order to interpret effectively and correctly the results obtained with these techniques. The underlying principles of NIR and IR spectroscopy are discussed in Chapter 1x of this book.

### 2.1 Principles of Chemical Imaging

Chemical, or hyper-spectral, imaging is based on the concept of image hyper-cubes that contain both spectral intensity and wavelength data for every 3-D image pixel; these are created as a result of spectral acquisition at every point of the microscopic chemical image. The intensity of a single pixel in such an image plotted as a function of the NIR or IR wavelength is in fact the standard NIR/IR spectrum for the selected pixel, and is usually represented as pseudo-color.

### 2.2 Principles of Fluorescence Correlation Spectroscopy/ Imaging

The presentation adopted here for the FCS principle closely follows a brief description recently developed by Eigen et al. (1). FCS involves a special case of fluctuation correlation techniques in which a laser light excitation induces fluorescence within a very small ($10^{-15}$ L = 1fL) volume of the sample solution whose fluorescence is auto-correlated over time. The volume element is defined by the laser beam excitation focused through a water- or oil-immersion microscope objective to an open, focal volume of ~ $10^{-15}$ L. The sample solution under investigation contains fluorescent molecule concentrations in the range from $10^{-9}$ to $10^{-12}$ M, and is limited only by detector sensitivity and available laser power. A non-invasive determination of single molecule dynamics can thus be made through fluctuation analysis that yields either chemical reaction constants or diffusion coefficients, depending on the system under consideration.

### 3.2. FT-IR and FT-NIR Microspectrometers

A microspectrometer is defined as a combination of a spectrometer and a microscope that has both spectroscopic and imaging capabilities. Such an instrument is capable, for example, of obtaining visible images of a sample using a CCD camera, and chemical images with an NIR detector. Chemical images are then employed for sophisticated quantitative analyses. The results reported in this chapter for soybean seeds and embryos were obtained with FT- IR and -NIR spectrometers made by the PerkinElmer Co. (Shelton, CT, USA). The FT-NIR (NTS model) spectrometer was equipped with an integrating sphere accessory for diffuse reflectance. The FT- IR or -NIR spectrometers were, respectively, attached to microscopes for the IR region (Spotlight 300) or NIR region (NIR Autoimage), as illustrated in Fig. 3.2.1 and Fig. 3.2.2. Each spectrometer has an internal desiccant compartment to remove the water vapor and the carbon dioxide from air that may interfere with the spectrum of a sample. Apart from the improved resolution and acquisition time, these instrument models, offer increased sensitivity and also allow the transfer of spectra to different instruments of similar design. The two microspectrometers are each equipped with two cassegrain imaging objectives and a third cassegrain before the NIR detector in order to improve focus and sensitivity, as shown in Fig. 3.2.3.

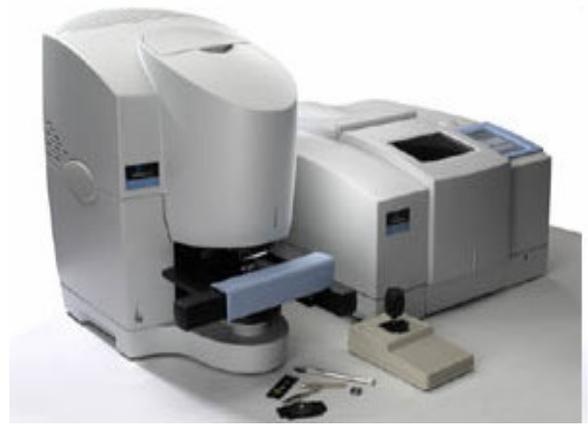

- ☐ Introduced in 2002 by the PerkinElmer Co. (Shelton, CT, USA) for high-resolution studies

- ☐ Employed for our mid-IR Microspectroscopy and Chemical Imaging investigations of thin sections of soybean seeds and embryos

**Fig. 3.2.1**. FT-IR Microspectrometer (Spotlight 300) introduced by the PerkinElmer Co.

- Introduced in 2002 by PerkinElmer Co. (Shelton, CT, USA) for high-resolution studies

- Employed for our NIR Microspectroscopy and Chemical Imaging investigations of soybean seeds and embryos

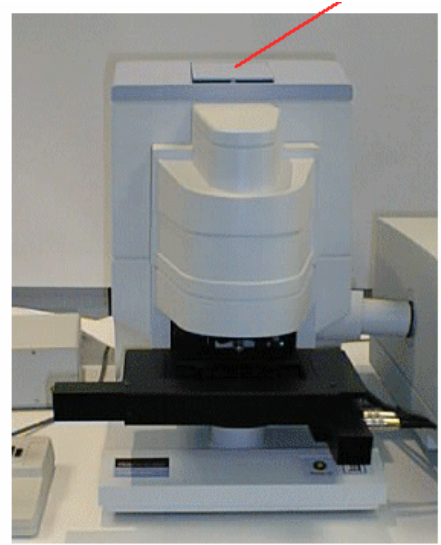

**Fig. 3.2.2.** FT-NIR Microspectrometer (AutoImage) made by the PerkinElmer Co.

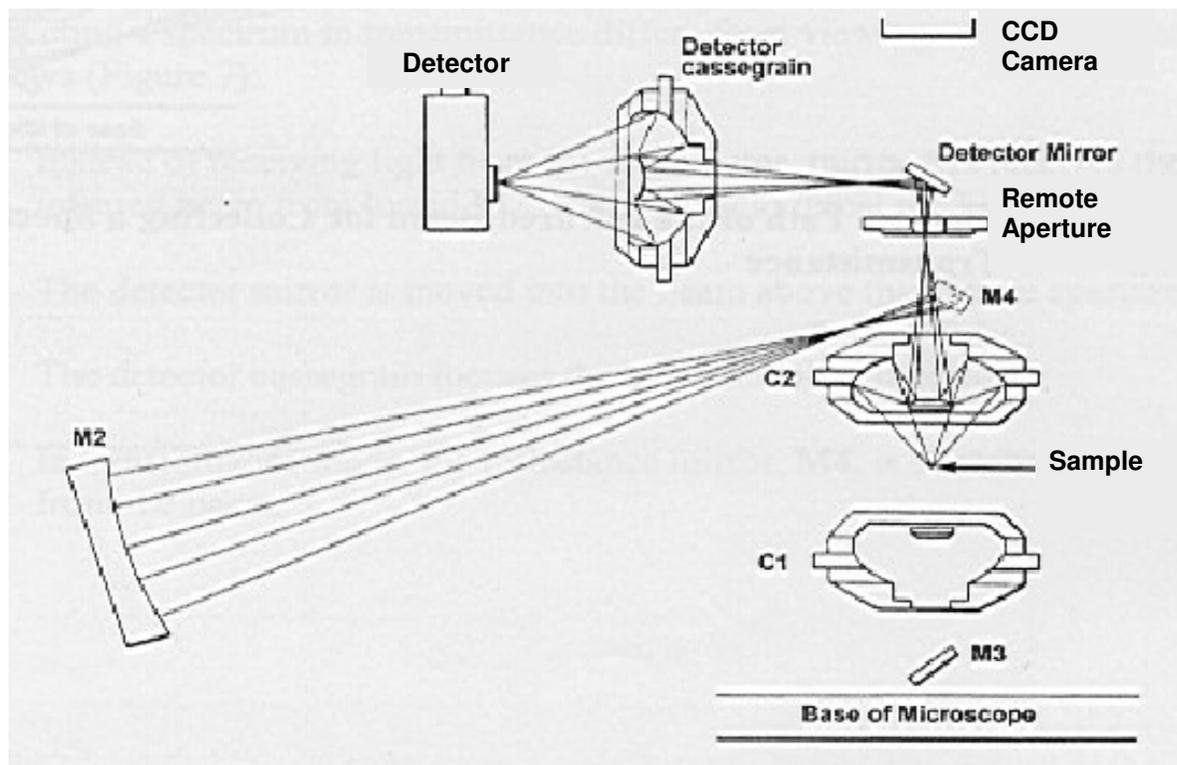

**Fig. 3.2.3**. A simplified diagram of the reflection mode of operation for the AutoImage FT-NIR Microspectrometer, manufactured by the PerkinElmer Co.

## 3.3. High-Resolution NMR Method for Oil Determination

The technique applied to obtain the oil content in soybean embryos was simple one-pulse, High-Resolution (HR) NMR (11). The HR-NMR technique was explained in Section 3.4 of Chapter 1x. A Varian U-400 NMR instrument was employed for oil measurements; the selected 90 deg pulse width was 19.4 μs and the $^1$H NMR signal absorption intensity was recorded with a 4 s recycling interval to avoid saturation.

## 3.4. Fluorescence Correlation Spectroscopy

This section presents submicron resolution imaging results that we obtained with two-photon NIR excitation of FCS. The FCS data was obtained in the Microscopy Suite of the Beckman Institute for Advanced Science and Technology at UIUC by employing two-photon NIR fluorescence excitation at 780 nm with a 180 fs, Ti: Sapphire pulsed laser, coupled to an FCS Alba™ spectrometer system (recently designed and manufactured by ISS Co., Urbana, Illinois). The configuration of an Alba$^{TM}$ spectrometer with an inverted microscope is shown in Fig. 3.4.1, and the optical detail path and the system components are presented in Fig. 3.4.2.

Multi-photon (MPE) NIR excitation of fluorophores--attached as labels to biopolymers like proteins and nucleic acids, or bound at specific biomembrane sites-- is one of the most attractive options in biological applications of laser scanning microscopy (12). Many of the serious problems encountered in spectroscopic measurements of living tissue, such as photodamage, light scattering and autofluorescence, can be reduced or even eliminated. FCS can therefore provide accurate *in vivo* and *in vitro* measurements of diffusion rates, "mobility" parameters, molecular concentrations, chemical kinetics, aggregation processes, labeled nucleic acid hybridization kinetics and fluorescence photophysics/photochemistry. Several photophysical properties of fluorophores that are required for quantitative analysis of FCS in tissues have already been reported (13). Molecular "mobilities" can be measured by FCS over a wide range of characteristic time constants from ~$10^{-3}$ to $10^3$ ms. At signal levels comparable to 1PE confocal microscopy, 2PE reduces photobleaching in spatially restricted cellular compartments, thereby preserving the long-term signal-to-noise during data acquisition (14). Furthermore, 3PE has been reported to eliminate DNA damage and photobleaching problems that may still be present in some 2PE experiments. Whereas both 1PE and 2PE alternatives are suitable for intracellular FCS observations on thin biological specimens, 2PE can substantially improve FCS signal quality in turbid samples, such as plant cell suspensions or deep cell layers within tissues.

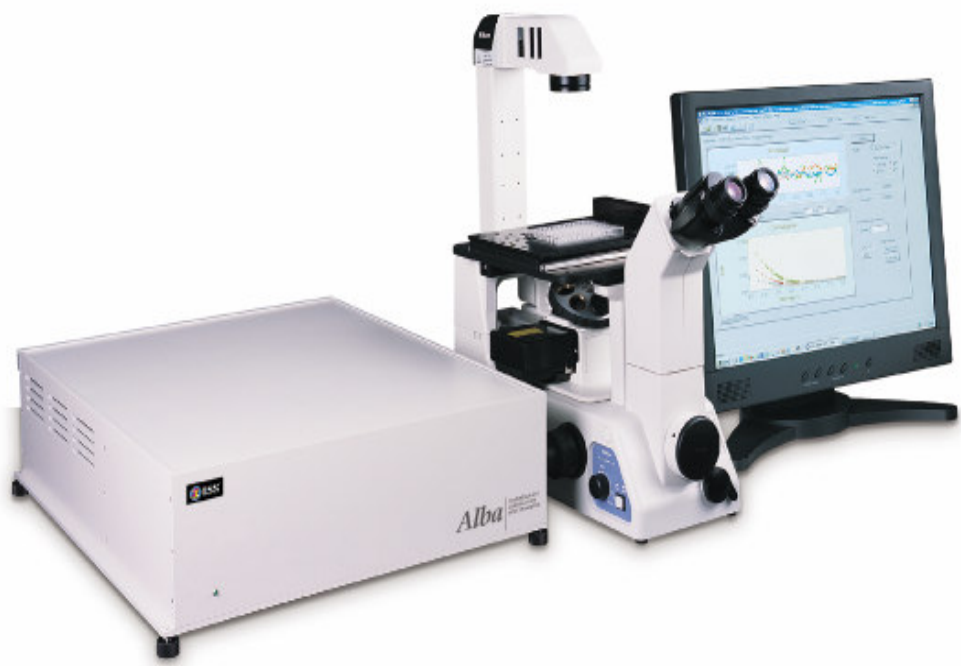

**Fig. 3.4.1.** The FCS Alba™ Microspectrometer System manufactured by ISS Co., (Urbana, Illinois, USA). The inverted, epi-fluorescence microscope shown in the figure is the Nikon TE-300 –special Model, that has available *both* a back illumination port and a left-hand side port. The PC employed for data acquisition, storage and processing is located behind the instrument, as is the laser illumination source (not visible in the figure).

**4.1. FT-IR and FT-NIR Chemical Imaging Tests**

A series of tests were carried out for both FT-NIR and FT-IR microspectrometers in order to compare both their imaging speed and microscopic resolution (15, 16). The results of such tests are presented, respectively, in Figures 4.1.1 and 4.1.2 for the Spotlight 300 model FT-IR, and in Figures 4.1.3 and 4.1.4 for the FT-NIR AutoImage microspectrometer. It is important to note the absence of spherical or chromatic aberrations in such images obtained with either the Spotlight 300 (FT-IR) or the AutoImage FT-NIR microspectrometers. In addition, one should also note that the spatial resolution increases dramatically to ~ 1 micron for the shorter NIR wavelengths, even with relatively thick samples, such as a 1 cm Zirconium single crystal (Fig. 4.1.3).

**3D Representation of Latex Beads**

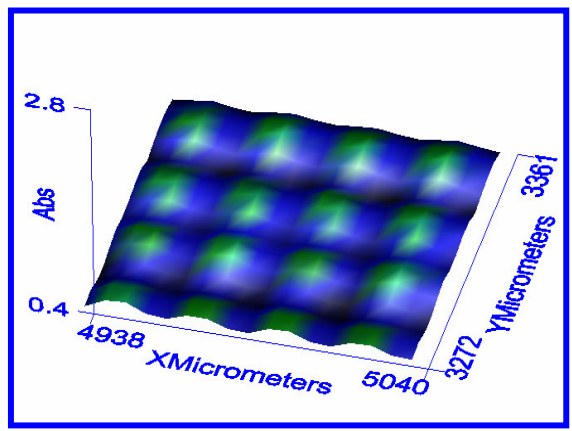

**Fig. 4.1.1.** FT-IR Single Wavenumber (761 cm$^{-1}$) Chemical Images of 6 micron diameter Latex Beads placed on an electron microscope grid. These FT-IR images were obtained with the Spotlight 300 IR-Microspectrometer using a total acquisition time of ~10 minutes, and demonstrate both the high imaging speed and the maximum microscopic resolution of this novel-design instrument.

=

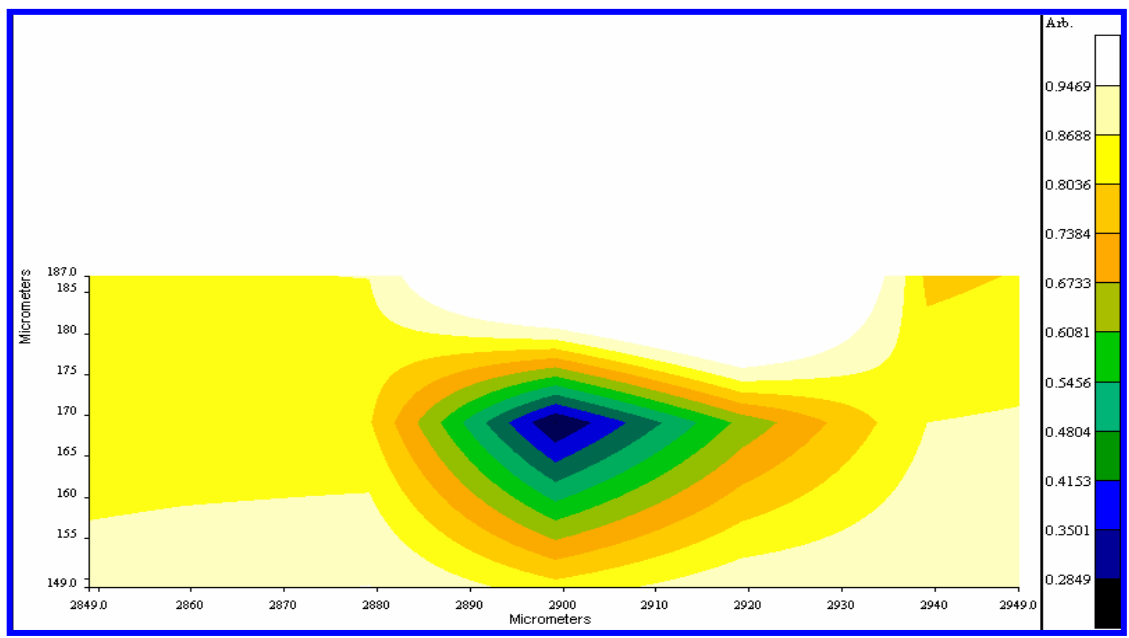

**Fig. 4.1.3**. Spatial Resolution Test: FT-NIR Reflection Mode Image of a 1 cm, Cubic Zirconium Single Crystal at a resolution of ~1 micron (Plot of the band ratio: 7253 to 5485 cm$^{-1}$).

## 6.2. FCCS Applications to DNA Hybridization, PCR and DNA Binding

In the bioanalytical and biochemical sciences FCS can be used to determine various thermodynamic and kinetic properties, such as association and dissociation constants of intermolecular reactions in solution (18, 19). Examples of this are specific hybridization and renaturation processes between complementary DNA or RNA strands, as well as antigene-antibody or receptor-ligand recognition. Although of significant functional relevance in biochemical systems, the hybridization mechanism of short oligonucleotide DNA primers to a native RNA target sequence could not be investigated in detail prior to the FCS/FCCS application to these problems. Most published models agree that the process can be divided into two steps: a reversible first initiating step, where few base pairs are formed, and a second irreversible phase described as a rapid zippering of the entire sequence. By competing with the internal binding mechanisms of the target molecule such as secondary structure formation, the rate-determining initial step is of crucial relevance for the entire binding process. Increased accessibility of binding sites, attributable to single-stranded open regions of the RNA structure at loops and bulges, can be quantified using kinetic measurements (20).

The measurement principle for nearly all our FCS/FCCS applications is based so far upon the change in diffusion characteristics when a small labeled reaction partner (e.g., a short nucleic acid probe) associates with a larger, unlabeled one (target DNA/RNA). The average diffusion time of the labeled molecules through the illuminated focal volume element is inversely related to the diffusion coefficient, and increases during the association process. By calibrating the diffusion characteristics of free and bound fluorescent partner, the binding fraction can be easily evaluated from the correlation curve for any time of the reaction. This principle has been employed to investigate and compare the hybridization efficiency of six labeled DNA oligonucleotides with different binding sites to an RNA target in a native secondary structure (20).

Hybridization kinetics were examined by binding six fluorescently labeled oligonucleotide probes of different sequence, length and binding sites to a 101-nucleotide-long native RNA target sequence with a known secondary structure (Fig. 6.1.1). The hybridization kinetics were monitored and quantitated by FCS, in order to investigate the overall reaction mechanism. In this "all-or-none" binding model, the expected second-order reaction was assumed to be irreversible. For nM concentrations and at temperatures around 40°C, the typical half-value reaction times for these systems are in the range of 30 to 60 min, and therefore the hybridization process could be easily followed by FCS diffusional analysis. At the measurement temperature of 40°C the probes are mostly denatured, whereas the target retains its native structure. The binding process could be directly monitored through diffusional FCS analysis, via the change in translational diffusion time of the labeled 17-mer to 37-mer oligonucleotide probes HS1 to HS6 upon specific hybridization with the larger RNA target (Fig. 6.1.1). The characteristic diffusion time through the laser-illuminated focal spot of the 0.5 µm-diameter objective increased from 0.13 to 0.20 ms for the free probe, and from 0.37 to 0.50 ms for the bound probe within 60 min. The increase in diffusion time from measurement to measurement over the 60 min could be followed on a PC monitor and varied strongly from probe to probe. HS6 showed the fastest association, while the reaction of HS2 could not be detected at all for the first 60 min. It has been shown above that FCS diffusional analysis provides an easy and comparably fast determination of the hybridization time course of reactions between complementary DNA/RNA strands in the concentration range from $10^{-10}$ to $10^{-8}$ M. Perturbation of the system is therefore not necessary, so the measurement can be carried out at thermal equilibrium. Thus, the FCS-based methodology also permits rapid screening for suitable anti-sense nucleic acids directed against important targets like HIV-1 RNA with low consumption of probes and target.

Because of the high sensitivity of FCS detection, the same principle can be exploited to simplify the diagnostics for extremely low concentrations of infectious agents like bacterial or viral DNA/RNA. By combining confocal FCS with biochemical amplification reactions like PCR or 3SR, the detection threshold of infectious RNA in human sera could be dropped to concentrations of $10^{-18}$ M (21, 22). The method is useful in that it allows for simple quantitation of initial infectious units in the observed samples. The isothermal Nucleic Acid Sequence-Based Amplification (NASBA) technique enables the detection of HIV-1 RNA in human blood-plasma (2). The threshold of detection is presently down to 100 initial RNA molecules per milliliter, and possibly much fewer in the future, by amplifying a short sequence of the RNA template (24; 25). The NASBA method was combined with FCS, thus allowing the online detection of the HIV-1 RNA molecules amplified by NASBA (22). The combination of FCS with the NASBA reaction was performed by introducing a fluorescently labeled DNA probe into the NASBA reaction mixture *at nanomolar concentrations*, hybridizing to a distinct sequence of the amplified RNA molecule. The specific hybridization and extension of this probe during the amplification reaction resulted in an increase of its diffusion time and was monitored online by FCS. Consequently, after having reached a critical concentration on the order of 0.1 to 1.0 nM (the threshold for FCS detection), the number of amplified RNA molecules could be determined as the reaction continued its course. Evaluation of the hybridization/extension kinetics allowed an estimation of the initial HIV-1 RNA concentration, which was present at the beginning of amplification. The value of the initial HIV-1 RNA number enables discrimination between positive and false-positive samples (caused, for instance, by carryover contamination). Plotted in a reciprocal manner, the slopes of the correlation curves in the HIV-positive samples drop because of the slowing down of diffusion after binding to the amplified target. This possibility of sharp discrimination is essential for all diagnostic methods using amplification systems (PCR as well as NASBA).

The quantitation of HIV-1 RNA in plasma by combining NASBA with FCS may be useful in assessing the efficacy of anti-HIV agents, *especially in the early infection stage when standard ELISA antibody tests often display negative results*. Furthermore, the combination of NASBA with FCS is not restricted only to the detection of HIV-1 RNA in plasma. Though HIV is presently a particularly common example of a viral infection, the *diagnosis of Hepatitis (both B and C) remains much more challenging*. On the other hand, the number of HIV, or HBV, infected subjects worldwide is increasing at an alarming rate, with up to 20% of the population in parts of Africa and Asia being infected with HBV. In contrast to HIV, HBV infection is not particularly restricted to the high-risk groups.

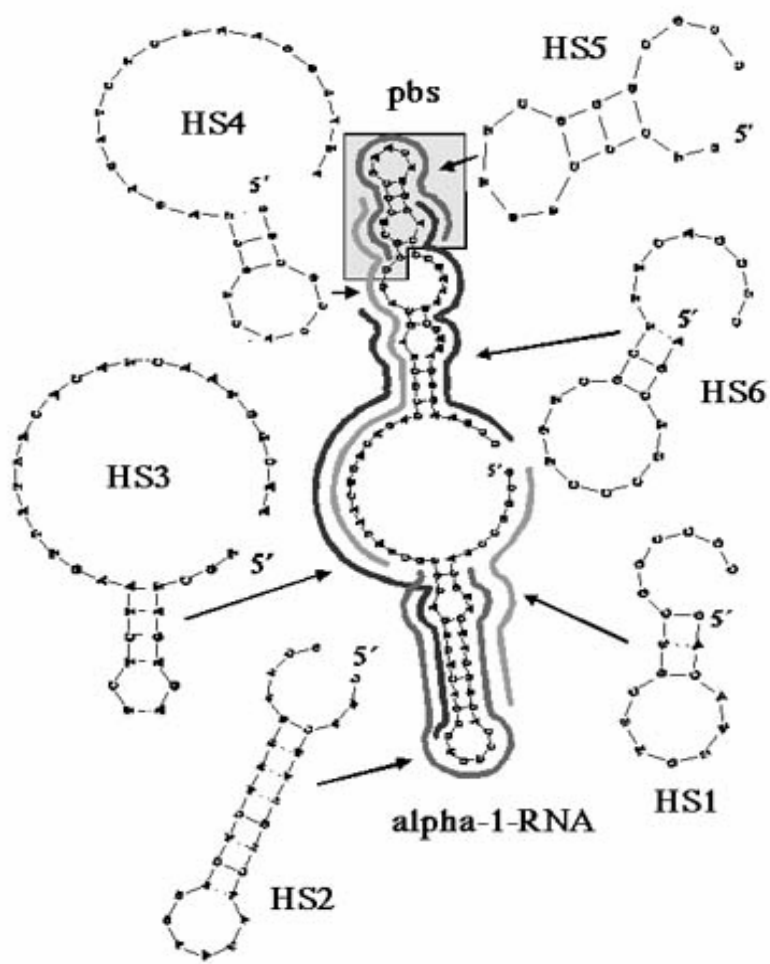

**Fig. 6.2.1.** Secondary structures and binding sites of the oligonucleotides HS1 to HS6 and the target RNA.

 FCCS Applications to DNA Hybridization, PCR and DNA Binding (Schwille (32).)

**CONCLUSIONS AND DISCUSSION**

Our results from high-resolution NMR analysis of oil with *nanoliter* precision in mutagenized somatic embryos strongly indicate that this novel methodology is practical for producing mature soybean embryos with increased oil content that would be of significant economical value. By comparison, the rate of useable mutants in whole soybean seeds has been reported to be as low as $10^{-4}$. Therefore, methodologies starting with whole mature soybean seeds have considerably higher cost and time requirements for the experimental selection process of mutant soybean lines with increased oil content, than our methodology that utilizes somatic embryos grown *in vitro*.

On the other hand, FT-NIR spectroscopy has major, practical advantages over other techniques (such as either low- or high-resolution NMR) for quantitative determination of oil, protein, moisture and perhaps even minor seed constituents. Such key advantages are: speed, accuracy, reproducibility, convenience (i.e., little or no sample preparation), and relatively low cost (in comparison with both pulsed and HR NMR). Furthermore, a significant advantage of the datasets/results obtained by adequately calibrated FT-NIR is the high internal consistency of the FT-NIR results for large numbers of normal, yellow-coat, soybean seed samples. These advantages are very important both for soybean breeding/selection programs and for wide-scale industrial applications of NIR composition analysis throughout the entire soybean distribution and processing chain. Major practical limitations of FT-NIR spectroscopy are the need for primary/reference methods and its lower resolution (compared with either FT-IR or HR-NMR).

Microscopic resolution testing of the Spotlight 300 model, FT-IR chemical imaging (array) microspectrometer with coated latex spheres and Mg-silicate particles yielded about 6 micron spatial resolution. The current, commercial instrumentation for FT-IR and FT-NIR Microspectroscopy/chemical imaging, and fluorescence microscopy is capable of *in vivo*, automated measurements and visualization of composition distribution in various cellular types and tissue systems.

Recent FT-NIR/IR developments and the combination of the FT-IR and FT-NIR spectroscopy with microscopy (i.e., Microspectroscopy) allow one to obtain microscopic, chemical images of soybeans and soybean embryos- both in reflection and transmission modes- in as little as 3 min at spectral resolutions up to ~8 $cm^{-1}$. The highest spatial resolution among the commercial FT microspectrometers investigated was close to 1 micron, and was obtained with the FT-NIR AutoImage model microspectrometer. In spite of its lower sensitivity (microgram vsus picogram, respectively), NMR Microscopy ('MRM') has also been reported to achieve 1 micron resolution under the most favorable conditions, at $^1$H resonance frequencies significantly lower than 1GHz (33; 34). At present, however, the typical resolution obtained by NMR Microscopy on seeds is on the order of 50 μm as in the case of castor bean imaging (33), or oil in the germ of wheat grains (35; 36). The resolution limit of NMR microscopy is limited by several factors (33) currently including the lower sensitivity compared with FT-NIR; nevertheless, technical improvements of NMR imaging techniques may be found that overcome such obstacles thus leading to submicron resolution. In this context, it is interesting that individual protein-bound water molecules could be observed in lysozyme by 2D NMR (37).

The latest developments indicate that the sensitivity range of FT-NIR microspectroscopy observations can be extended to the *femtogram* level, with submicron spatial resolution. Such FT-NIR/IR microspectroscopy instrumentation developments are potentially very important for those agricultural and food biotechnology --as well as biomedical and pharmacological applications-- that require rapid and sensitive analyses, such as the screening of high-content microarrays in Genomics and Proteomics research. Novel, two-photon NIR excitation fluorescence correlation microspectroscopy results were here reported with submicron resolution for concentrated suspensions of plant cells and thylakoid membranes. With advanced super-resolution microscopy designs, a further, tenfold resolution increase is attainable, in principle, along the optical (z) axis of the microspectrometer. Especially promising are current developments employing multi-photon NIR excitation that could lead, for example, to novel cancer prevention methodology and the early detection of cancers using NIR-excited fluorescence. Other related developments are the applications of Fluorescence Cross-Correlation Spectroscopy detection to monitoring DNA hybridization kinetics, DNA binding, ligand-receptor interactions and HIV-HBV testing.

Very detailed, automated chemical analyses of oils/fats and phytochemicals (e.g., isoflavones in cell cultures) are now also becoming possible by FT-NIR microspectroscopy of single cells, either *in vitro* or *in vivo*. Such rapid analyses have potentially important applications in food safety, agricultural biotechnology, medical research, pharmacology and clinical diagnosis.

**FIGURE CAPTIONS**

**Fig. 2.2.1.** An experimental setup for a Single-Photon, Confocal Fluorescence Correlation Spectroscopy, according to Eigen et al. (1).

**Fig. 2.2.2.** Autocorrelation function, and Tau plotted as a histogram and as a function of time. (Adapted from Winkler et al.(2))

**Fig. 2.2.3.** Fluorescence Intensity Fluctuations caused by various Dynamic Processes. (Adapted from Winkler et al. (2))

**Fig. 2.3.1.** Illustration of the FCCS Principle. (Adapted from Winkler et al. (2)).

**Fig. 3.2.1**. FT-IR Microspectrometer (Spotlight 300) introduced by the PerkinElmer Co.

**Fig. 3.2.2.** FT-NIR Microspectrometer (AutoImage) made by the PerkinElmer Co.

**Fig. 3.2.3**. A simplified diagram of the reflection mode of operation for the AutoImage FT-NIR Microspectrometer, manufactured by the PerkinElmer Co.

**Fig. 3.4.1.** The FCS Alba™ Microspectrometer System manufactured by ISS Co., (Urbana, Illinois, USA). The inverted, epi-fluorescence microscope shown in the figure is the Nikon TE-300 –special Model, that has available *both* a back illumination port and a left-hand side port. The PC employed for data acquisition, storage and processing is located behind the instrument, as is the laser illumination source (not visible in the figure).

**Fig. 3.4.2.** Diagram of an FCS Spectrometer Coupled to an Inverted Epi-fluorescence Microscope. (Adapted from Eigen *et al.* (1)).

**Fig. 4.1.1.** FT-IR Single Wavenumber (761 cm$^{-1}$) Chemical Images of 6 micron diameter Latex Beads placed on an electron microscope grid. These FT-IR images were obtained with the Spotlight 300 IR-Microspectrometer using a total acquisition time of ~10 minutes, and demonstrate both the high imaging speed and the maximum microscopic resolution of this novel-design instrument.

**Fig. 4.1.2.** FT-IR Image Resolution Test: 3D Surface Projection View of FT-IR Image of 10μ Diameter Spheres Obtained with the FT-IR Spotlight 300 Microspectrometer.

**Fig. 4.1.3**. Spatial Resolution Test: FT-NIR Reflection Mode Image of a 1 cm, Cubic Zirconium Single Crystal at a resolution of ~1 micron (Plot of the band ratio: 7253 to 5485 cm$^{-1}$).

**Fig. 4.1.4.** FT-NIR Microimaging of 25-Micron Micro-arrays.

**Fig. 4.2.1**. FT-IR Reflectance Chemical Images Compared with the Visible Reflectance Image (middle picture) of a black-coated soybean obtained with a PerkinElmer Spotlight 300 Chemical Imaging/FPA Microspectrometer. The soybean region labeled "Y" shows a zone where the black coat was removed, thus unveiling the yellow soybean interior which has a markedly different IR absorption spectrum from that of the black coat region.

**Fig. 4.2.2.** FT-NIR Chemical Image of Oil Distribution in a Mature Soybean Embryo Section.

**Fig. 4.2.3.** Visible Image of a Soybean Somatic Embryo: sample at 9.6% moisture, seen at 60X magnification.

**Fig. 4.2.4**. Transmission IR Chemical Image of a Live Soybean Somatic Embryo. Shown in red is the high intensity of the protein amide II absorption band.

**Fig. 4.2.5.** FT-NIR Reflectance Chemical Image of a Red Coat Bean (*Vigna angularis*) and a Section without the Red Coat (left, below diagonal).

**Fig. 4.2.6.** FT-NIR Chemical Imaging by Difference Spectroscopy (CIDS): Red Coat Spectra Minus the Yellow, Under-Coat Section Spectra.

**Fig. 4.3.1.** The soybean oil standard plot for 400 MHz $^1$H NMR measurements on the Varian U400. The probe was a Nalorac 5 mm QUAD for high-resolution liquids $^1$H NMR.

**Fig. 4.3.2.** $^1$H NMR Spectrum of the somatic embryogenic culture of a soybean sample (10A10). The $^1$H NMR spectra were collected with a Varian U400 spectrometer and a Nalorac 5 mm $^1$H Quad probe tuned at a 400 MHz resonance frequency in an external magnetic field of 9.4T (with FT of 500 NMR transients). Several broad absorption bands are present between 0 and 3 ppm, whereas several $^1$H NMR sharp peaks of oil are located between 3.1 ppm and 4.3 ppm. The $^1$H NMR peak of water (HDO) is located at 4.7 ppm.

**Fig. 4.3.3.** TEM Micrograph of a Suspension of Soybean Somatic Embryos in Culture (after treatment with 1 mM Ethyl Methane Sulfonate). Oil droplets (visible at arrows) are distributed inside embryos, and were more frequent in mutant soybean embryos with the higher oil contents.

**Fig. 6.1.1.** Fitted, Single-Photon, Auto-Correlation Plots: **A**. 1 and 2 nM Solutions of the Fluorescent Dye Rhodamine 110G obtained with 1PE at 488 nm; **B.** 10 nM Solution of the Fluorescent Dye Rhodamine 110G; **C.** FCS Computer Simulation Results.

**Fig. 6.1.2**. Fluorescence Photon-counting Histogram corresponding to the Correlation in Fig. 3.4.1B.

**Fig. 6.1.3.** Test of FCS for Two-Photon 780 nm NIR Excitation with 0.25 micron fluorescently labeled spheres. The continuous line shows the regression fitting obtained with the Vista software program from ISS (Copyright 2003).